\documentclass[aps,pra,superscriptaddress,twocolumn]{revtex4}

\usepackage{bm}
\usepackage{graphicx}
\usepackage{amsmath}

\DeclareMathOperator{\sgn}{sgn}
\DeclareMathOperator{\mmax}{max}
\DeclareRobustCommand\openzero{\leavevmode\hbox{0\kern-.55em0}}

\begin{document}

\renewcommand{\H}{{\cal{H}}}
\newcommand{\F}{{\cal{F}}}
\newcommand{\bS}{{\bm{S}}}

\newcommand {\mean}[1]{\langle #1 \rangle}
\newcommand {\meantwo}[1]{\langle #1 \rangle}
\newcommand{\parall}{\uparrow\uparrow}
\newcommand{\antiparall}{\uparrow\downarrow}
\newcommand\lavg{\left\langle}
\newcommand\ravg{\right\rangle}
\newcommand\ket[1]{|{#1}\rangle}
\newcommand\bra[1]{\langle{#1}|}
\newcommand\bket[1]{\big|#1\big\rangle}
\newcommand\bbra[1]{\big\langle#1\big|}

\title {Optimal dynamics for quantum-state and entanglement transfer
through homogeneous quantum wires}

\author{L. Banchi}
\affiliation{Dipartimento di Fisica, Universit\`a di Firenze,
             Via G. Sansone 1, I-50019 Sesto Fiorentino (FI), Italy}

\author{T.~J.~G.  Apollaro}
\affiliation{Istituto dei Sistemi Complessi, Consiglio Nazionale delle Ricerche,
             via Madonna del Piano 10, I-50019 Sesto Fiorentino (FI), Italy}

\author{A. Cuccoli}
\affiliation{Dipartimento di Fisica, Universit\`a di Firenze,
             Via G. Sansone 1, I-50019 Sesto Fiorentino (FI), Italy}

\affiliation{CNISM - Consorzio Nazionale Interuniversitario per le
Scienze Fisiche della Materia}

\author{R. Vaia}
\affiliation{Istituto dei Sistemi Complessi, Consiglio Nazionale delle Ricerche,
             via Madonna del Piano 10, I-50019 Sesto Fiorentino (FI), Italy}

\author{P. Verrucchi}
\affiliation{Istituto dei Sistemi Complessi, Consiglio Nazionale delle Ricerche,
             via Madonna del Piano 10, I-50019 Sesto Fiorentino (FI), Italy}
\affiliation{Dipartimento di Fisica, Universit\`a di Firenze,
             Via G. Sansone 1, I-50019 Sesto Fiorentino (FI), Italy}
\affiliation{INFN Sezione di Firenze, via G.Sansone 1,
             I-50019 Sesto Fiorentino (FI), Italy}
\date{\today}

\begin{abstract}
It is shown that effective quantum-state and entanglement transfer
can be obtained by inducing a coherent dynamics in quantum wires with
homogeneous intrawire interactions. This goal is accomplished by
tuning the coupling between the wire endpoints and the two qubits
there attached, to an optimal value. A general procedure to determine
such value is devised, and scaling laws between the optimal coupling
and the length of the wire are found. The procedure is implemented in
the case of a wire consisting of a spin-$\frac12$ $XY$ chain: results
for the time dependence of the quantities which characterize
quantum-state and entanglement transfer are found of extremely good
quality and almost independent of the wire length. The present
approach does not require {\em ad hoc} engineering of the intrawire
interactions nor a specific initial pulse shaping, and can be applied
to a vast class of quantum channels.
\end{abstract}

\maketitle

One of the most commonly requested conditions in quantum
communication and computation protocols is that two distant parties,
typically Alice and Bob, share a couple of entangled qubits. When the
physical objects encoding the qubits can travel, as in the case of
optical photons, the above goal can be accomplished by creating the
entangled couple in a limited region of space and then letting the
qubits fly where necessary. On the other hand, when qubits are
realized via intrinsically localized physical objects, as in the case
of $S\,{=}\,\frac12$ spins or atomic systems, a different strategy
must be adopted (see for instance Ref.~\onlinecite{Bose2007} and
references therein). One such strategy is the following: first, two
neighboring qubits (A and $\rm{A}'$) are prepared in an entangled
state, by means of a short-range interaction; then, the mixed state
of one of the two qubits (say A) is transferred to a third distant
qubit via a quantum channel. If state-transfer is perfect the
procedure results in a pair of distant entangled qubits A$'$ and B,
as requested.

Aim of this paper is to set a general framework where such strategy
can be successful. In particular, ({\em i}) we define a procedure for
controlling such dynamics, and hence the quality of the transmission
process, by specific operational settings; ({\em ii}) we show that
the quality of the quantum-state and entanglement transfer is not
substantially affected by the length of the wire; ({\em iii}) we
apply the procedure to the spin-$\frac12$ $XY$ chain and show that
high-quality quantum-state and entanglement transfer are obtained.

Let us first recall that for the strategy depicted above to make
sense, one has to equip oneself with a quantum channel capable of
transferring mixed states. How to obtain such a channel is the
problem to which many authors have proposed different
solutions~\cite{Bose2007,Bose2003,OsborneL2004,%
ChristandlDDEKL2005,Haselgrove2005,CamposVenutiBR2007,DiFrancoPK2008},
some based on the idea of engineering the channel itself, by the
specific design of its internal interactions, others on that of
intervening on the initialization process, by preparing the wire in a
configuration found to serve the purpose. In both cases, a severe
external action on the physical system is required.

Here a different point of view is adopted: the purpose is that of
devising conditions for an optimal dynamics to occur, where by
`optimal dynamics' we mean a time evolution of the state of the
channel such that the mixed state of A is transferred with high
fidelity through the channel to a distant party B. The main guideline
is the fact that excitations characterized by a linear dispersion
play a crucial role in determining effective transfer of quantum
states. Since this idea has been put
forward~\cite{OsborneL2004,YangSS2008,CubittC2008}, it had to face
the unfortunate evidence that a quantum channel with a linear
dispersion relation over the whole Brillouin zone is not only
difficult to design~\cite{ChristandlDDEKL2005}, but perhaps just a
chimera to realize, so far. On the other hand, most physical systems
are characterized by excitations whose dispersion relation has zones
of linearity, typically near inflection points: if we were able to
induce a dynamical evolution of the channel essentially ruled by
excitations belonging to such zones, a coherent propagation should
result, and an effective transfer of quantum states consequently
obtained. Let us hence focus on how to induce such dynamical
evolution.

\begin{figure}[t]
\includegraphics[width=70mm,angle=0]{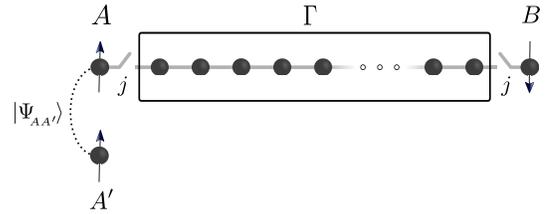}
\caption{The endpoints of a quantum wire $\Gamma$ are coupled to the
qubits A and B, via a switchable interaction $j$; A can be entangled
with an external qubit A$'$.}
\label{f.qwire}
\end{figure}

We consider (see Fig.~\ref{f.qwire}) two separated parties, Alice and
Bob, each owning a qubit, A and B, respectively. Each qubit may be
subjected to a local interaction,
$h_{_{\rm{Q}}}{\cal{H}}_{_{\rm{A}}}$ and
$h_{_{\rm{Q}}}{\cal{H}}_{_{\rm{B}}}$, respectively, with
$h_{_{\rm{Q}}}$ possibly tunable. The quantum channel is realized by
an extended physical system $\Gamma$, hereafter called {\em quantum
wire}, which is here assumed to be made of $N$ interacting particles
on a discrete lattice. The internal dynamics of the wire is ruled by
the Hamiltonian ${\cal{H}}_{_\Gamma}$. The endpoints of the wire can
be put in contact with the qubit A (B) via a switchable interaction
$j{\cal{H}}_{_{\rm{A}\Gamma}}$ ($j{\cal{H}}_{_{\rm{B}\Gamma}})$. The
overall system has mirror symmetry and the Hamiltonian which rules
its dynamics is $ {\cal{H}}={\cal{H}}_{_\Gamma} +
j({\cal{H}}_{_{\rm{A}\Gamma}}+{\cal{H}}_{_{\rm{B}\Gamma}}) +
h_{_{\rm{Q}}}({\cal{H}}_{_{\rm{A}}}+{\cal{H}}_{_{\rm{B}}})~. $ Before
the process starts A, B and $\Gamma$ do not interact with each other;
the wire is in its ground state $\ket{\Omega_{_\Gamma}}$, while Alice
and Bob prepare their qubits in the initial states
$\rho_{_{\rm{A}}}(0)$ and $\rho_{_{\rm{B}}}(0)$, respectively. At
time $t\,{=}\,0$ the interaction between each qubit and the wire is
switched on.

We assume that: ({\em i}) ${\cal{H}}$ can be written (either exactly
or via a reasonable approximation) as a quadratic form of $N+2$
local, either fermionic or bosonic, operators
$\{\eta_i,\eta_i^\dagger\}$, where the index
$i\,{=}\,0,\,\dots,\,N{+}1$ labels the sites of the lattice; ({\em
ii}) the bulk of the wire is translation symmetric. Condition ({\em
i}) guarantees the existence of a unitary transformation
$\{\eta_i,\eta_i^\dagger\}\to\{\eta_k,\eta_k^\dagger\}$, which
diagonalizes the total Hamiltonian,
$\H=\sum_k\omega_k\eta_k^\dagger\eta_k$. Condition ({\em ii}) entails
that the above transformation is close to a Fourier transform when
$i$ corresponds to sites in the bulk of the wire, so that $k$ can be
approximately considered as a quasi-momentum. For the sake of
clarity, let us consider the qubits A and B as initially prepared in
the pure states $\ket{\alpha}$ and $\ket{\beta}$, respectively; the
overall system at time $t\,{=}\,0$ is then described by
$\ket{\Psi_0}=\ket{\alpha}\ket{\Omega_{_\Gamma}}\ket{\beta}$, which
generally is a non-equilibrium state. Its dynamics is ruled by
quasi-particle excitations with density in $k$-space
$n(k)=\bra{\Psi_0}\eta_k^\dagger\eta_k\ket{\Psi_0}$, which, in their
turn, evolve according to $\H$. If the dispersion relation were
linear, $\omega_k\,{\simeq}\,\omega_0\,{+}\,vk$, the bulk dynamics
would consist in a coherent wavepacket traveling from A to B (and
vice versa) at velocity $v$ along the wire. Under our further
assumption of mirror symmetry, such wavepacket would perfectly
rebuild the initial state of A on the qubit B, after a time
$t\,{\simeq}\,N/v$.

\begin{figure}[t]
\includegraphics[height=60mm,angle=90]{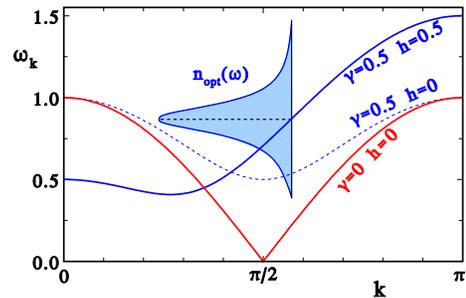}
\caption
{Dispersion relation of the spin-$\frac12$ $XY$ chain, for different
interaction parameters. The bell-shaped curve edging the shaded area
is the optimal density of excitation vs. $\omega$, as obtained for
$\gamma=\gamma_{0,N}=0.5$, and $h=0.5$. }
\label{f.omegak}
\end{figure}

However, dispersion relations of real interacting systems are
typically non linear, and display a much more complex dependence on the
wavevector $k$ (see for instance Fig.~\ref{f.omegak}).
On the other hand, if we cannot deal with a most unusual linear dispersion,
yet we can act on the distribution
$n(k)$ by varying the coupling between the
qubits and the wire, or other parameters entering the Hamiltonian of the qubits, or
their initial states $\ket{\alpha}$ and $\ket{\beta}$.

Suppose then that we can adjust the distribution $n(k)$ to make it
peaked around an inflection point $k_0$ of the dispersion relation, thus
setting the most relevant excitations for the dynamics, as those whose
energies can be written as
\begin{equation}
 \omega_k\simeq\omega_{k_0}+v(k{-}k_0)+\frac{2a}3(k{-}k_0)^3 ~,
\label{e.omega3}
\end{equation}
where the main dispersive term is retained. Further suppose that we
can set the width of the distribution $n(k)$: be $1/(4\sigma^2)$ its
variance. It can be shown~\cite{MiyagiN1979} that, under the above
conditions, the dynamics of the wire is still described by a
traveling spacial wavepacket, centered at
$x(t)=\big(v+a\,\sigma^{-2}\big)\,t$, with a variance that increases
from the initial value $\sigma^2(0)=\sigma^2$ as
$\sigma^2(t)=\sigma^2+{a^2t^2}/(2\sigma^4)$~.

We hence state that the optimal dynamics is obtained by choosing
$\sigma^2$ so as to minimize $\sigma(t)$ at the arrival time
$t\,{\simeq}\,N/v$, leading to
\begin{equation}
 \sigma_{\rm{opt}}= \sqrt[3]{|a|t}
      \simeq \big(|a|N/v\big)^{1/3}~.
\label{e.sigmaopt}
\end{equation}

As stated above, the form of the dispersion relation is essentially
designed by the wire Hamiltonian, the position of the peak of $n(k)$
can be set by acting on $h_{_{\rm{Q}}}$, while $\sigma$ is expected
to depend on the coupling between the qubits and the wire, whose
energy scale is $j$.

Based on the above reasoning, we define the following procedure for
inducing an optimal dynamics: Fix the Hamiltonian parameters of the
wire so as to grant the existence of a wide enough interval in $k$
with almost linear dispersion~\cite{footnote1}, as described by
Eq.(\ref{e.omega3}); tune $h_{_{\rm{Q}}}$ so as to peak $n(k)$ at an
inflection point of $\omega_k$; then set $\sigma$ to its optimal
value Eq.~(\ref{e.sigmaopt}) by varying $j$. The value of the
coupling corresponding to $\sigma_{\rm{opt}}$ will be hereafter
referred to as the {\em optimal coupling}, $j_{\rm{opt}}$. It is
worth mentioning that the optimization of $\sigma$ can also be
obtained by creating a finite-size excitation in space as the
starting
state~\cite{PaganelliGP2009,OsborneL2004,Haselgrove2005,BishopUWB2010};
however, this strategy results in a much less versatile procedure.
The weak dependence on $N$ embodied in Eq.~(\ref{e.sigmaopt}) is
expected to give small losses for transmissions over long wires. In
fact, the actual time $t_N$ needed for the packet to travel along the
wire, implicitly defined by $x(t_N)=N$, is found to depend on $N$
according to
$v\,t_N\,{\simeq}\,N-\frac{\sgn{a}}2\,\sigma_{\rm{opt}}$. It is of
absolute relevance that
$\sigma(t_N)\simeq\sqrt{3/2}\,\sigma_{\rm{opt}}$, i.e., the final
optimal packet is just about 22\,\% wider than at start,
irrespectively of the wire length.

Let us now describe a specific implementation of the procedure
described above. We consider a system defined on a one-dimensional
discrete lattice of $N{+}2$ sites, labeled by the index
$i=0,~1,\,\dots,~N{+}1$. The qubits A and B sit at sites $0$ and
$N{+}1$, respectively. The wire is physically realized by a chain of
$N$ interacting $S\,{=}\,\frac12$ spins, taking up sites from $1$ to
$N$. Neighboring spins interact via a Heisenberg Hamiltonian of the
$XY$ type, and are possibly subjected to an external magnetic field.
The exchange interaction and the magnetic field are assumed
homogeneous along the wire. The total Hamiltonian is
\begin{eqnarray}
&\,&\!\!\!\!\H{=}
-\sum_{i=1}^{N-1}(1{+}\gamma)S_i^x S_{i+1}^x+(1{-}\gamma)S_i^y S_{i+1}^y
{-}h\sum_{i=1^N} S_i^z\nonumber\\
&\,&\!\!\!\!{-}j\sum_{i=0,N}\Big[(1{+}\gamma_i)S_i^x
S_{i+1}^x+(1{-}\gamma_i)S_i^y S_{i+1}^y
\Big]\nonumber\\
&\,&\!\!\!\!{-}h_{_{\rm{Q}}}(S_0^z+S_{N+1}^z),
\label{e.Htot}
\end{eqnarray}
where the exchange energy for the intrawire interaction has been
taken as the reference energy scale and hence set to unity. Mirror
symmetry implies $\gamma_N\,{\equiv}\,\gamma_0$. An isotropic
exchange interaction in Eq.~\ref{e.Htot},
$\gamma\,{=}\,\gamma_0\,{=}\,0$, defines the so called $XX$ model. A
Jordan-Wigner transformation, casts the Hamiltonian (\ref{e.Htot})
into a quadratic form of $N{+}2$ interacting fermionic operators,
$\{\eta_i,\eta^\dagger_i\}$, each defined on a site of the lattice. A
further Bogolubov transformation diagonalizes the Hamiltonian,
$\H\,{=}\,\sum_k\omega_k\eta_k^\dagger\eta_k\,{+}\,E_0$, where $E_0$
is the ground state energy, whereas $\eta^\dagger_k~(\eta_k)$ are
fermionic operators which create (annihilate) excitations of energy
$\omega_k$~\cite {LiebSM1961}.

The optimization procedure described above begins with the analysis
of the dispersion relation in the infinite chain limit,
$\omega_k\,{=}\,[(h{-}\cos{k})^2+\gamma^2\sin^2k]^{1/2}$, displayed
in Fig.~\ref{f.omegak}. One can easily spot the existence of more or
less wide regions of linearity in the neighborhood of the inflection
point(s) $k_0$, where Eq.~(\ref{e.omega3}) holds. By tuning
$h_{_{\rm{Q}}}\,{\simeq}\,\omega_{k_0}$, the peak of $n(k)$ is made
to sit at $k_0$. The optimal coupling $j_{\rm{opt}}$ is then
numerically determined so as to fulfill Eq.~(\ref{e.sigmaopt}). In
fact, the dispersion relation strongly depends on the parameters of
the wire Hamiltonian, $\gamma$ and $h$: in particular, the region of
linear dispersion sensibly shrinks as the anisotropy $\gamma$
increases, which might make the wire to be useless; however, the
linear region can be extended again by increasing the field $h$:
therefore, one can act on the latter parameter so as to fulfil the
conditions for optimal dynamics. For example, in the extreme case of
the Ising chain ($\gamma\,{=}\,1$) for $h\,{=}\,0$ the dispersion
relation becomes flat and does not allow for propagation; however, a
wide linear region can yet be obtained by applying a finite $h$ on
$\Gamma$.

\begin{figure}
\includegraphics[width=80mm,angle=0]{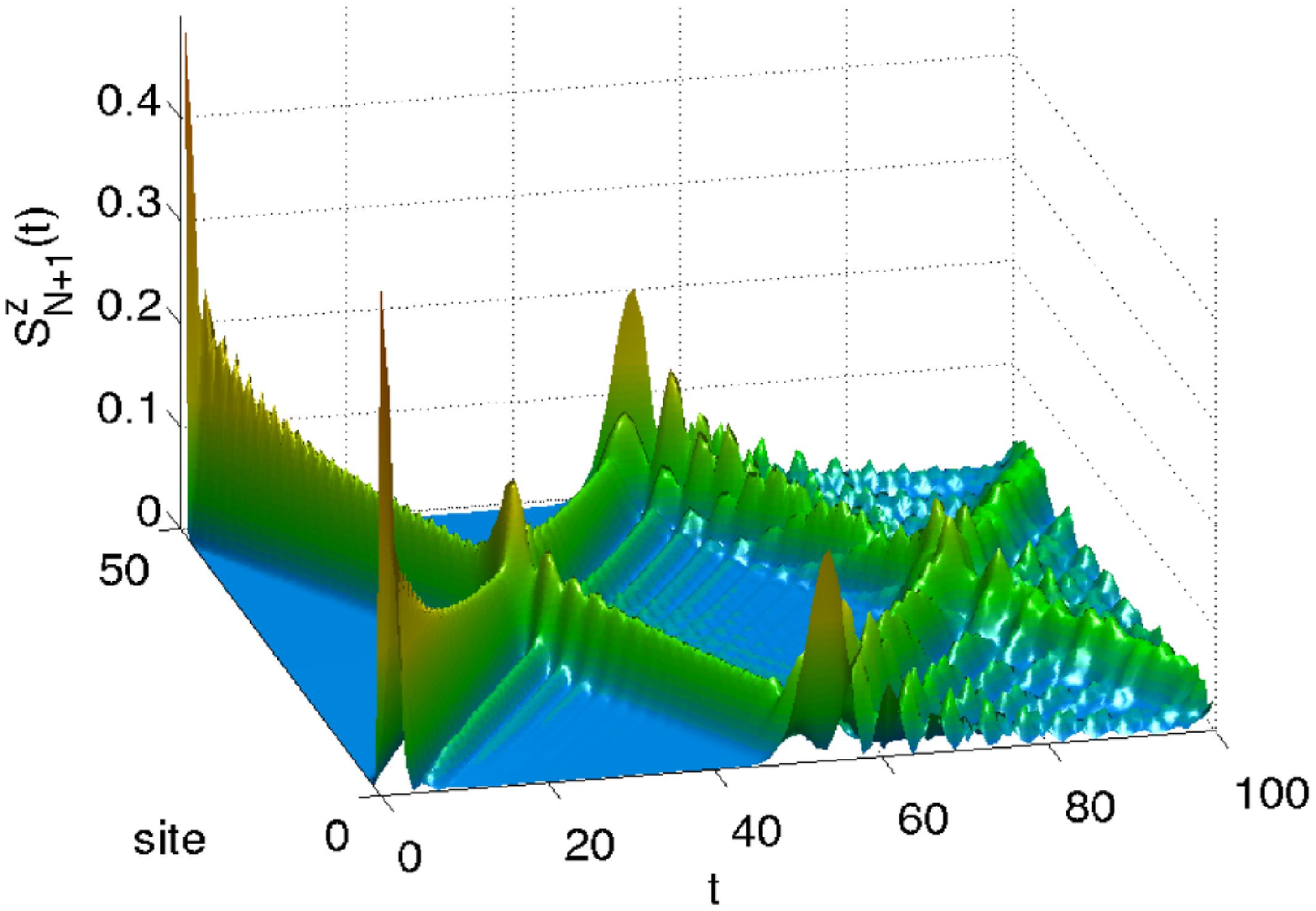}
\includegraphics[width=80mm,angle=0]{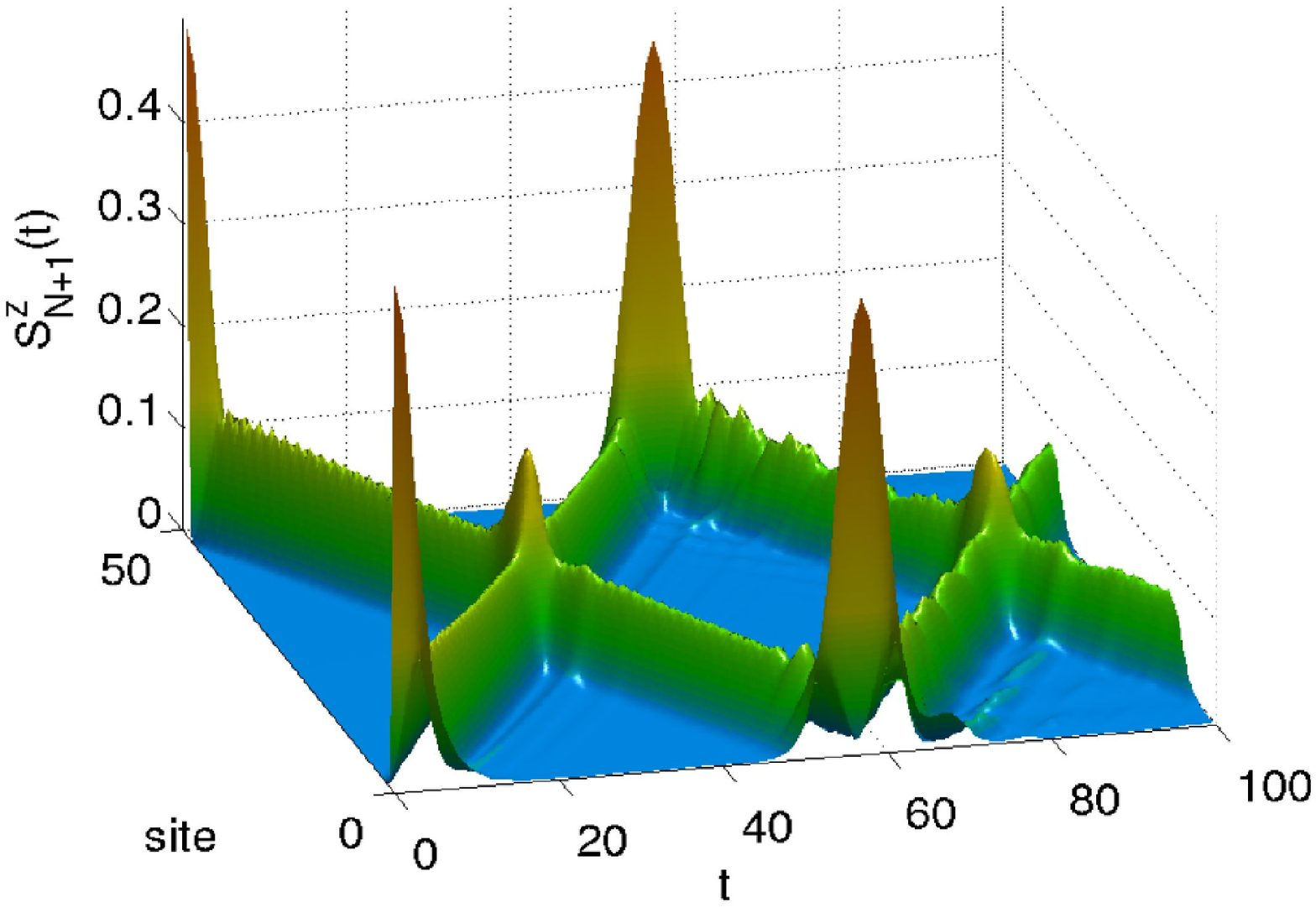}
\caption{Time evolution of the on-site $z$-magnetization for
$N\,{=}\,50$ and an initial state
$\ket{\uparrow}\ket{\Omega_{_\Gamma}}\ket{\uparrow}$; the qubit-wire
coupling is $j\,{=}\,1$ (top) and $j\,{=}\,j_{\rm{opt}}\,{=}\,0.58$
(bottom).}
\label{f.sz}
\end{figure}

Let us now give a flash upon the resulting dynamics of the overall
system. To this aim, we set $\gamma\,{=}\,\gamma_0\,{=}\,0$ ($XX$
model) and $h\,{=}\,0$, as this choice makes more analytical
expressions available, which in turn allows for a more detailed
analysis. The inflection point $k_0\,{=}\,\pi/2$ corresponds to
$\omega_{k_0}\,{=}\,0$ and we hence set $h_{_{\rm{Q}}}\,{=}\,0$.

In Fig.~\ref{f.sz} we show the time evolution of the magnetization
parallel to the quantization axis all along the wire,
$\mean{S^z_i(t)}, i\,{=}\,0,\,1,\dots,\,N,\,N{+}1$, for $N\,{=}\,50$
and initial states of the qubits
$\ket{\alpha}\,{=}\,\ket{\beta}\,{=}\,\ket{\uparrow}$. The upper
panel corresponds to a generic value of the coupling, $j\,{=}\,1$.
The lower panel is for $j\,{=}\,j_{\rm{opt}}=0.58$: this value is
determined exactly by exploiting analytical expressions holding for
the $XX$ model~\cite{WojcikLKGGB2005}. The difference from the
previous panel is striking: indeed, the dynamics of the wire is
essentially ruled by the nondispersive propagation of wavepackets,
and results in $\mean{S^z_{_{\rm{B}}}(t)}$ being, at the arrival time
$t\simeq N$, an almost perfect reproduction of the initial
magnetization $\mean{S^z_{_{\rm{A}}}(0)}$ of the qubit A. These
results do also confirm that our prescription for the above coherent
propagation to occur, i.e., for the determination of $j_{\rm{opt}}$,
is correct.

But does this peculiar dynamical evolution also affect the quality of
quantum-state transfer? To answer this question, we now deal with the
time evolution of quantities which are used to monitor such quality.
In particular, we analyze the quantum-state transfer process in terms
of the fidelity between an initial state $\ket{\alpha}$ of A and the
evolved state $\rho_{_{\rm{B}}}(t)$ of B,
$\F_{_{\rm{AB}}}(t)\,{=}\,\bra{\alpha}\rho_{_{\rm{B}}}\!(t)\ket{\alpha}$,
where initially
$\rho_{_{\rm{B}}}\!(0)\,{=}\,\ket\uparrow\bra\uparrow$. As for the
entanglement, we refer to the time evolution of the concurrence
between A$'$ and B, $\mathcal{C}_{_{\rm{BA}'}}\!(t)
\equiv\mathcal{C}\big(\rho_{_{\rm{BA}'}}\!(t)\big)$~\cite{footnote2}.
In the case of the $XX$ model we can also evaluate the {\em minimum}
fidelity over all possibile $\ket{\alpha}$, hereafter indicated by
$\F_{_{\rm{AB}}}^{\text{min}}(t)$, which allows us to put forward
further conclusions about the quality of the entanglement
transmission from AA$'$ to BA$'$. Indeed, this can be done using the
following lower bound~\cite{Nielsen,KnillL97} for the fidelity of
entanglement:
$\F\big(\ket{\psi_{_{\rm{AA}'}}}\bra{\psi_{_{\rm{AA}'}}},
\rho_{_{\rm{BA}'}}(t)\big)\ge\frac32\F^{\rm{min}}_{_{\rm{AB}}}(t)-\frac{1}{2}$.
The entangled state $\ket{\Psi_{_{\rm{AA}'}}}$ can be hereafter taken
as any of the Bell states, since a local operation on A$'$ does not
change the dynamics of the concurrence.

The above quantities essentially depend on the time evolution of the
magnetizations of B, which we have determined as follows: The
components $S_{N+1}^\alpha$ have been written in terms of the
fermionic operators $\{\eta_k,\eta_k^\dagger\}$; their
time-dependence, which simply follows from
${\cal{H}}=\sum_k\omega_k\eta_k\eta_k^\dagger$, has been made
explicit, so as to obtain, by transforming back via inverse Bogolubov
and Jordan-Wigner transformation, the Heisenberg representation of
the spin operators $S_{N+1}^\alpha(t)$. Their expectation values,
i.e., the required magnetizations, have been finally derived by
expanding $S_{N+1}^\alpha(t)$ in terms of the spin operators of the
extremal qubits and of fermionic operators relative to the wire,
i.e., defined by the diagonalization of the sole
${\cal{H}}_{_\Gamma}$. This approach allows us to devise a numerical
procedure for determining, for any model belonging to the class
described by the Hamiltonian (\ref{e.Htot}), the Kraus
operators~\cite{Nielsen} $M_\mu(t)$, in terms of which we get
$\rho_{_{\rm{B}}}(t)={\sum_\mu}M_\mu(t)\rho_{_{\rm{A}}}(0)M^\dagger_\mu(t)$
and $\rho_{_{\rm{BA}'}}(t)=\sum_\mu\big[M_\mu(t)\otimes1\big]
\rho_{_{\rm{AA}'}}(0)\big[M_\mu^\dagger(t)\otimes1\big]$,
i.e., the necessary tools for our analysis.

\begin{figure}
\includegraphics[width=80mm,angle=0]{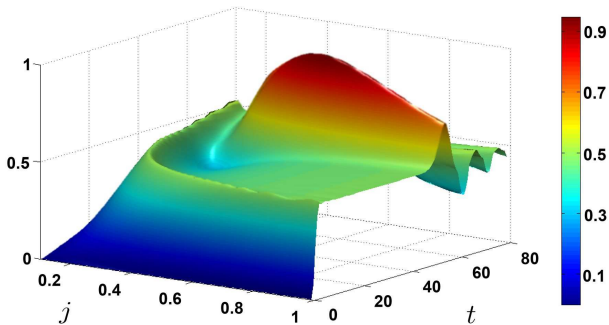}
\caption{Time evolution of the minimum fidelity
$\F_{_{\rm{AB}}}^{\text{min}}(t)$ vs $j$, for a $N\,{=}\,50$,
$\gamma\,{=}\,\gamma_0\,{=}\,\gamma_N\,{=}\,0$ ($XX$ chain), and
$h\,{=}\,h_{_{\rm{Q}}}\,{=}\,0$.}
\label{f.mfidxx}
\end{figure}

We first consider the fidelity $\F_{_{\rm{AB}}}(t)$ as defined above.
In the case of the $XX$ chain, we put ourselves in the worst possible
case and evaluate $\F_{_{\rm{AB}}}^{\text{min}}(t)$: a high value of
such quantity ensures a very good transfer of {\em any} initial
state, modulo a local operation. In Fig.~\ref{f.mfidxx} we show
$\F_{_{\rm{AB}}}^{\text{min}}(t)$ for the same model as in
Fig.~\ref{f.sz} as a function of the coupling $j$. A minimum fidelity
which is just slightly below unity is observed for that same value of
the coupling, determined by our procedure, which gives rise to the
dynamics shown in the lower panel of Fig.~\ref{f.sz}, i.e.,
$j\,{=}\,j_{\rm{opt}}\,{\simeq}\,0.58$.

We underline that such a high value of the minimum fidelity, which is
confirmed also for $N$ as large as $500$, does also imply an optimal
transmission of entanglement. It is worth noticing that the peak of
$\F_{_{\rm{AB}}}^{\text{min}}(t)$ occurs simultaneously with that of
$\mean{S^z_{_{\rm{B}}}(t)}$, i.e., at $t\simeq N$, a feature that can
be analytically proven in the $XX$ case.

The consistence of the above picture allows us to move forward to the
analysis of the entanglement transfer, and also in the more general
case of the $XY$ model. We set $\gamma\,{=}\,\gamma_0\,{=}\,0.5$ and
$h\,{=}\,0.5$ and apply our procedure: We first locate the inflection
point of the dispersion relation in the thermodynamic limit at
$k_0\,{\simeq}\,1.795$ and tune $h_{_{\rm{Q}}}\,{\simeq}\,0.85$. This
latter value slightly differs from $\omega_{k_0}$ as determined in
the thermodynamic limit, due to finite-size and boundary effects.
Finally, Eq.~({\ref{e.sigmaopt}) allows us to numerically determine
$j_{\rm{opt}}\,{=}\,0.49, 0.39, 0.34$ for $N\,{=}\,50, 250, 500$,
respectively.

\begin{figure}[t]
\includegraphics[height=60mm]{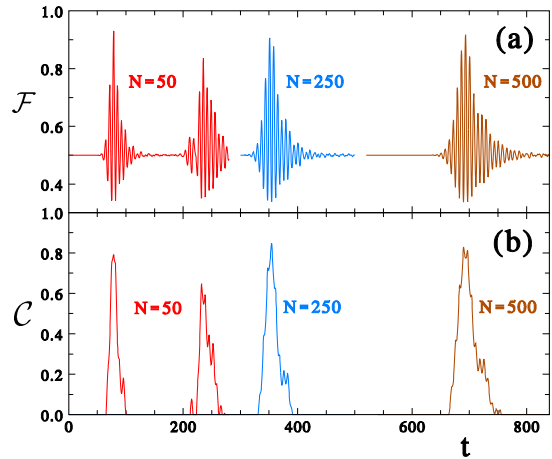}
\caption{Time evolution of
$\bar{\cal{F}}_{\rm{AB}}$ (upper panel) and
$\mathcal{C}_{_{\rm{BA}'}}(t)$ (lower panel), for different $N$,
$\gamma\,{=}\,\gamma_0\,{=}\,0.5$, $h\,{=}\,0.5$,
$h_{_{\rm{Q}}}\,{=}\,0.85$, and $j_{\rm{opt}}\,{=}\,0.49, 0.39, 0.34$
for $N\,{=}\,50, 250, 500$, respectively. The second signal observed
in both panels is the first echo for $N\,{=}\,50$.}
\label{f.XYfavc}
\end{figure}

Despite the dynamics in the $XY$ case being complicated by the
existence of several energy scales, we manage to get a very good
state transfer also for large $N$, as testified by the fidelity,
averaged over all possible initial states of A,
$\bar{\cal{F}}_{_{\rm{AB}}}$ shown in the upper panel of
Fig.\ref{f.XYfavc}. The oscillations observed are due to the
precession induced by the magnetic field $h_{_{\rm{Q}}}$ acting on
the qubit B. Such precession causes only a phase change and does not
affect the entanglement transfer, as shown below. The entanglement
between B and A$'$ is monitored by the time evolution of the
concurrence $\mathcal{C}_{_{\rm{BA}'}}\!(t)$, shown in the lower
panel of Fig.~\ref{f.XYfavc} for different values of $N$: For each
$N$, this is characterized by an extremely well defined peak, of
height $> 0.8$ even for chains as long as $500$ sites. This peak
occurs simultaneously with the maximum of
$\bar{\cal{F}}_{_{\rm{AB}}}$ and has a finite but small width.

The second signal observed in both panels is the first echo for
$N\,{=}\,50$, which is seen to be still considerably intense and well
localized: this testifies of a long-lasting quasi non dispersive
dynamics of the wire. Other echoes occur at times which are too large
for being shown in the figure, but their structure confirm this
statement.

In conclusion, based on a general picture of the dynamical evolution
of quantum wavepackets, we have devised a procedure for inducing an
optimal dynamics in a quantum wire. The procedure leads to the
determination of an optimal coupling between the wire and the two
qubits attached at its endpoints, meanwhile giving indications about
the best setting of other, possibly tunable, parameters of the
system. For the procedure to apply, few very simple conditions must
be fulfilled, and there is no need for a specific design neither of
the wire, nor of its initial state. By implementing our approach to
the spin-$\frac12$ $XY$ chain, we have obtained extremely good
quantum-state and entanglement transfer. The time scales over which
such transfer is found to occur is considerably shorter as compared
with previous results concerning quantum-state transmission over
Heisenberg
models~\cite{Bose2007,WojcikLKGGB2005,PlastinaA2007,GualdiKMT2008,DoroninZ2010}.
Moreover, the quality of the state and entanglement transfer that we
obtain very weakly deteriorate as the length of the wire increases.

\medskip

PV gratefully thanks Dr.~N. Gidopoulos for useful discussions, and
the ISIS centre of the Science and Technology Facilities Council (UK)
for the kind hospitality. PV also acknowledges financial support from
the italian CNR under the  "Short-term mobility 2010" funding scheme.


\begin{thebibliography}{00}
\bibitem{footnote1}
If the dispersion relation $\omega_k$ does not display a wide
interval of linearity, the wire is unlikely to behave as a good
quantum channel, at least by the mechanism considered here.
\end{thebibliography}
\end{document}